\documentclass[%
 aip,
 jmp,%
 amsmath,amssymb,
 reprint,%
]{revtex4-1}

\usepackage{graphicx}
\usepackage{dcolumn}
\usepackage{bm}
\usepackage{color}

\usepackage[utf8]{inputenc}
\usepackage{graphicx}
\usepackage{amsmath}

\newcommand{\figref}[1]{Fig.~\ref{#1}}
\newcommand{\figrefstartsentence}[1]{Figure~\ref{#1}}
\newcommand{\eqnref}[1]{Eq.~\ref{#1}}

\newcommand\micron{\mbox{$\mu$m}}%

\begin{document}

\title[Tile-and-trim micro-resonator array fabrication optimized for high multiplexing factors]{Tile-and-trim micro-resonator array fabrication optimized for high multiplexing factors}

\author{Christopher M. McKenney}
 \affiliation{National Institute of Standards and Technology}
\author{Jason E. Austermann}
 \affiliation{National Institute of Standards and Technology}
\author{Jim Beall}
 \affiliation{National Institute of Standards and Technology}
\author{Bradley Dober}
 \affiliation{National Institute of Standards and Technology}
\author{Shannon M. Duff}
 \affiliation{National Institute of Standards and Technology}
\author{Jiansong Gao}
 \affiliation{National Institute of Standards and Technology}
\author{Gene C. Hilton}
 \affiliation{National Institute of Standards and Technology}
\author{Johannes Hubmayr}
 \affiliation{National Institute of Standards and Technology}
\author{Dale Li}
 \affiliation{SLAC National Accelerator Laboratory}
\author{Joel N. Ullom}
 \affiliation{National Institute of Standards and Technology}
\author{Jeff Van Lanen}
 \affiliation{National Institute of Standards and Technology}
\author{Michael R. Vissers}
 \affiliation{National Institute of Standards and Technology}

\date{\today}
\begin{abstract}
We present a superconducting micro-resonator array fabrication method that is scalable, reconfigurable, and has been optimized for high multiplexing factors.  
The method uses uniformly sized tiles patterned on stepper photolithography reticles as the building blocks of an array.  
We demonstrate this technique on a 101-element microwave kinetic inductance detector (MKID) array made from a titanium-nitride superconducting film.  
Characterization reveals 1.5\% maximum fractional frequency spacing deviations caused primarily by material parameters that vary smoothly across the wafer.
However, local deviations exhibit a Gaussian distribution in fractional frequency spacing with a standard deviation of $2.7 \times 10^{-3}$. 
We exploit this finding to increase the yield of the BLAST-TNG 250~\micron\ production wafer by placing resonators in the array close in both physical and frequency space.  
This array consists of 1836 polarization-sensitive MKIDs wired in three multiplexing groups. We present the array design and show that the achieved yield is consistent with our model of frequency collisions and is comparable to what has been achieved in other low temperature detector technologies.

\end{abstract}

\maketitle

\section{Introduction}

Large arrays of low temperature detectors are needed to meet mapping speed requirements of future submillimeter astronomy and millimeter cosmology missions\cite{S4Ref,CCATP}.  
Increased detector counts necessitate improved fabrication throughput with high detector yield.  
Readout of larger detector arrays also requires greater multiplexing factors - the number of detectors per cryogenic wiring and amplifier chain.  
Arrays based on superconducting transition edge sensor (TES) bolometers and SQUID multiplexers that achieve multiplexing factors of $\sim$ 70 have enabled focal planes of $\sim$ 10,000 pixels\cite{SCUBA2,SPT3G} with typical end-to-end detector yields of 65-85\%\cite{ACT:1,SCUBA2,BICEP3:1}. 

An alternative approach is the Microwave Kinetic Inductance Detector (MKID)\cite{MKIDCaltech}, which utilizes the complex surface impedance of a superconductor\cite{MattisBardeen} to produce a resonator with a quality factor and resonant frequency sensitive to absorbed radiation.  
High quality factors ($> 10^{4}$) and resonator frequency definition through photolithography enable large-pixel-count focal planes with high multiplexing factors via frequency division multiplexing at RF or microwave frequencies.  
On-sky cameras have demonstrated multiplexing factors of $\sim 200 - 500$ with end-to-end detector yields of 70-80\%\cite{ARCONS:1,NIKA2}.

Scaling MKIDs to larger arrays presents two primary challenges that are addressed in this paper.  
Traditional stepper-lithography techniques excel at step-and-repeat patterns of identical cells.
MKID arrays however require lithographic variations per cell in order to achieve unique resonant frequencies.
In addition, improving MKID array yield requires reducing the number of resonator frequency collisions, which occur when the bandwidths of two resonators strongly overlap.  
The collision rate is determined by the designed fractional frequency spacing $\Delta = \left(f_{n+1} - f_{n}\right) / f_{n}$, the actual scatter of $\Delta$ in the fabricated wafer, the quality factor $Q$, and the minimum number of bandwidths $\chi$ allowed between resonators.
If the realized scatter in $\Delta$ is Gaussian distributed with standard deviation $\sigma$, the probability $P$ that a resonator survives all possible collisions with $n$ other resonators in a large array is given by \cite{2017MKIDFreqTrimming}
\begin{align}
    \label{Eq:TheorySurvive}
	P = \Pi_{n=1}^{n=\inf} \left( 1 - \frac{\text{erf}\left(\frac{n \Delta + \chi / Q}{\sqrt{2} \sigma}  \right) - \text{erf}\left(\frac{n \Delta - \chi / Q}{\sqrt{2} \sigma}  \right)}{2} \right)^{2}.
\end{align}
$P$ also gives the actual array yield due to frequency collisions, which increases with wider frequency spacing (larger $\Delta$), higher $Q$, lower $\sigma$, and smaller $\chi$.  To enable high multiplexing density (equivalent to low $\Delta$), we strive to minimize $\sigma$.

Here we present a tile-and-trim layout and fabrication method that utilizes stepped lithography to realize arbitrary MKID focal plane geometries and is optimized for frequency spacing uniformity.  
This approach builds on previous MKID work, which demonstrated $<$1\% variation in the superconducting transition temperature ($T_c$) across a 76.2~mm wafer by use of a TiN/Ti/TiN multilayer film \cite{TiN_Multilayer}.

This paper is organized as follows.  
In Section II we present the tile-and-trim fabrication method as well as the design and fabrication details of a test wafer used to demonstrate the method.   
Section III describes the measurements. 
In Section IV we present results and discussion of the achieved frequency spacing uniformity and the physical detector parameters that give rise to non-uniformity across the array.
In Section V we demonstrate the efficacy of the approach through the design and fabrication of the 250~$\mu$m focal plane for BLAST-TNG\cite{BLASTTNG}, a balloon-borne submillimeter telescope.  
We conclude in Section VI.

\section{Design and Fabrication Approach}

\begin{figure*}[ht!]
  \centering
    \includegraphics[width=\textwidth]{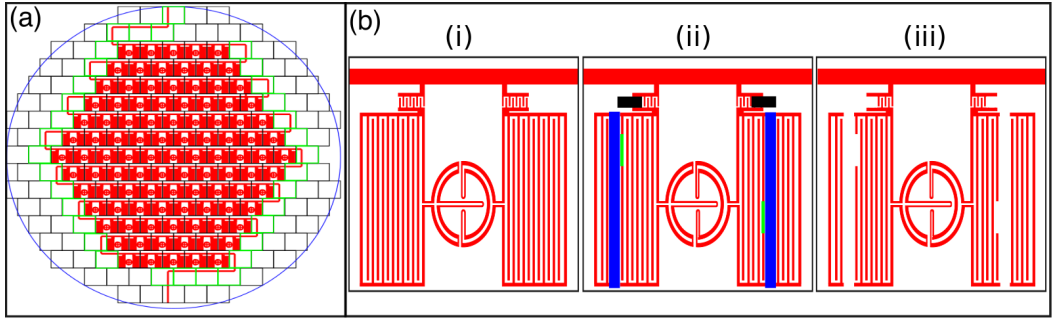}
      \caption{Tile-and-trim micro-resonator array fabrication method applied to an MKID test wafer.
      (a) We grid a wafer into rectangular cells that match a desired hexagonal-close-packed geometry.  
      In this example, the detector pitch is 5.0~mm and is intended to align with a monolithic feedhorn array.
      Each cell contains either an MKID pixel (majority of the wafer, and defined with multiple lithography exposures) or a piece of the microstrip transmission line (cells outlined in green and defined with a single exposure). 
      (b) The MKID cell definition uses a series of trimming steps to create unique $LC$ resonances. 
      (i) The base pixel is printed in each cell with one exposure.  
      The pixel consists of two orthogonally oriented absorbers, which also serve as the inductive element in an $LC$ resonator.  
      Interdigitated capacitors (IDCs) are used in both the resonators and to couple to the microwave feedline.    
      (ii) Subsequent exposure of trimmer tiles on each base pixel cut each IDC to length, setting the resonant frequencies.  
      Frequency keying is achieved by use of both a coarse trimmer tile (blue) that removes multiple IDC fingers and a fine trimmer tile (green) that removes a section of the last IDC finger.  
      As coupling to the feedline is frequency dependent, an additional tile (black) trims the feedline IDC to set the coupling quality factor $Q_c$.    
      (iii) Post etching, the remaining structure has two $LC$ resonators each with the desired number of IDC fingers.  Remnant IDC features also remain but have negligible effect on the detector performance.  
      }
  \label{Fig:CartoonProcess}
\end{figure*}

The tile-and-trim fabrication technique is an efficient method to produce micro-resonator arrays.  
For example given a base pixel design, an array containing 1000s of MKIDs with unique resonant frequencies may be laid out in a matter of minutes.  
We have demonstrated the fabrication of such an array in less than a week by use of a minimal set of photolithography reticles.  
The tile-and-trim method uses uniformly sized tiles defined on stepper reticles as the building blocks of an array.  
The procedure is illustrated in \figref{Fig:CartoonProcess}.
A base pixel is printed on a grid, and the resonant frequency is defined by use of `trimmer' reticle tiles stepped across the grid in a second pass.  These trimmer tiles remove sections of the capacitive elements in the base pixel. 
Exposing each wafer cell to the same base tile ensures a high level of lithographic uniformity throughout the array.  
Separate reticle tiles are then stepped to wire the microwave feedline to the edge of the wafer for easy interfacing. The tiling process allows the geometry to be easily modified if the pixel or microwave port locations change.

The base pixel design is the same as is used in BLAST-TNG \cite{BLASTTNG}, which is intended for feedhorn optical coupling and has previously been described \cite{NISTMKID:1,Dober2016}. 
In brief, the pixel contains two orthogonally polarized absorbers, which are optically sensitive to Cooper pair breaking radiation. 
Each absorber also acts as an independent superconducting inductor at microwave readout frequencies, which in parallel with an interdigitated capacitor (IDC) determines the resonant $LC$ frequency.  
The $3 \times 550 \; \micron^{2}$ absorber / inductor geometry is identical for each pixel, while the $LC$ frequency is varied via the number of IDC fingers using the tile-and-trim approach.  
The base $2 \; \micron$ width-and-separation IDC consists of 176 fingers, which are $1.2 \; \text{mm}$ long.  

Wafers are fabricated by first depositing a 60~nm thick TiN/Ti/TiN multilayer\cite{TiN_Multilayer}, which has a a superconducting transition temperature $T_{c}  \approx 1.7 \; \text{K}$.  
After the tile-and-trim lithography process the multilayer is dry-etched in a fluorine-chemistry Inductively Coupled Plasma Reactive Ion Etch (ICP-RIE).  
The CHF$_{3}$/Ar/O$_{2}$ etch is optimized to produce high internal quality factor ($Q_{i}$ $>$ $10^{5}$)  while also achieving a relatively slow etch rate into the silicon substrate.  
The ratio of the etch rates of the TiN multilayer and Si are $\sim 1:1$.  

\begin{figure*}[ht]
  \centering
    \includegraphics[width=\textwidth]{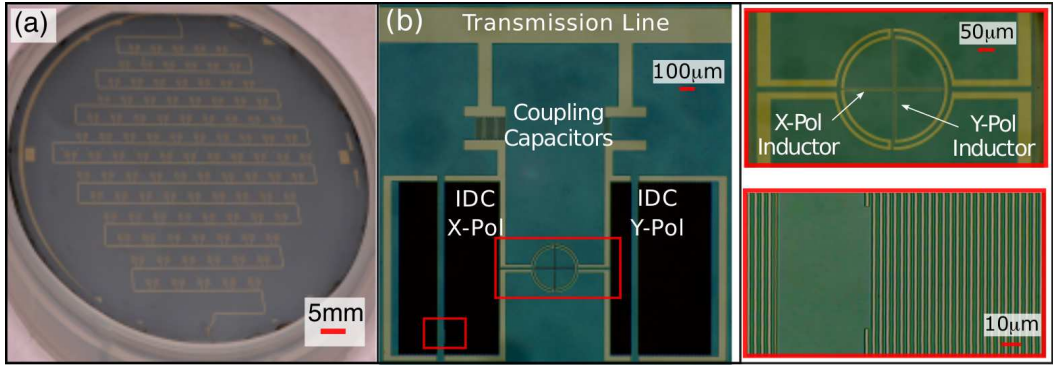}
      \caption{Photographs of realized test array.  
      (a) Full array of 101 pixels.  The tiling scheme matches that shown in \figref{Fig:CartoonProcess}(a).
      (b) Micrograph of one dual-polarization MKID pixel with labeled components.  
      (c) Zoom-in on the X-Polarization (X-Pol) and Y-Polarization (Y-Pol) inductor/absorbers, which are identical for all pixels.  The waveguide output of a feedhorn array is co-centric with the cross feature.  
      (d) Zoom-in on the trimmed IDC.  The traces in the left hand side of the micrograph comprise the remnant IDC structure.  
      The fingers on the right belong to the MKID IDC.  
      The fine trim structure can be seen on the left most finger of this IDC.       
    }
  \label{Fig:DevicePhotos}
\end{figure*}

Using the described layout and fabrication, we produced a $76.2 \; \text{mm}$ diameter test array with 101 pixels tiled by use of $5.00 \; \text{mm} \times 4.33 \; \text{mm}$ rectangular cells, which matches a 5~mm pitch hexagonal-close-packed geometry.  
\figrefstartsentence{Fig:CartoonProcess} shows the test wafer layout, and photographs of the finished array with labeled components are shown in \figref{Fig:DevicePhotos}.
The coarse, fine, and feedline coupling trimmer tiles (\figref{Fig:CartoonProcess}(b)) have dimensions $50 \times 1300 \; \micron^{2}$, $2 \times 100 \; \micron^{2}$, and $325 \times 160 \; \micron^{2}$, respectively.

For each of the two MKIDs in the pixel, the trimming procedure leaves behind two remnant, resonant structures.
The coarse trim produces an IDC in parallel with the MKID, and the fine trim leaves a partial electrically floating finger.  
Simulations show that the presence of these structures cause $<10^{-5}$ fractional frequency shift in the MKID resonant frequency and are thus negligible.

The test wafer contains two groups of resonators to explore different aspects of frequency spacing uniformity.    
Pixels belonging to group 1 consist of 90 resonators with designed frequencies spanning 560-612~MHz, $\Delta~=~1.0\times10^{-3}$, in a serpentine pattern of ascending frequency.
The trimming scheme is applied to the X-Polarization (X-Pol) IDC in the resonator, whereas the Y-Polarization (Y-Pol) MKIDs are designed to have resonant frequencies $> 2 \; \text{GHz}$ so as to be out of band of the X-Pol resonators being probed.  
This group targets wafer-scale frequency spacing uniformity.

Pixels in group 2 consist of 11 identically designed resonators with resonant frequency $f=700 \; \text{MHz}$ in the center row of the wafer.  
Pixels in this group were individually diced out post full-wafer characterization for independent measurement of film properties that affect the frequency spacing uniformity.
As such the tiles for the center row pixels include a $3 \; \text{mm} \times 3 \; \micron$ test wire used to determine $T_{c}$ and sheet resistance ($R_{s}$) of the TiN/Ti/TiN multilayer.  

\section{Measurement Description}

The test wafer was packaged in a 100 mm copper carrier equipped with SMA ports and mounted to the cold stage of a two-stage adiabatic demagnetization refrigerator (ADR) capable of cooling to $\sim 50 \; \text{mK}$.  
The microwave output of the detector package was connected to a cryogenic amplifier with a noise temperature $T_{n} \approx 5 \; \text{K}$.  
We used a vector network analyzer (VNA) to measure the microwave transmission of the array at $100 \; \text{mK}$ and fit the resulting resonator responses by use of the resonant circle approach\cite{GaoThesis} to extract the resonant frequencies. 

After array level measurements were performed, the 11 center pixels were individually diced and mounted in an ADR.  
Four wire measurements on the test wire structure were used to determine $T_c$ (measured at DC) and $R_s$ just above the superconducting transition.  
Separately the etch depth of the Si substrate adjacent to the IDC fingers was measured by use of a stylus profilometer.  
The resonant frequency of each identically designed center row MKID was independently measured in a dilution refrigerator.

\section{Test Array Results and Discussion}

\begin{figure}[ht!]
  \centering
    \includegraphics{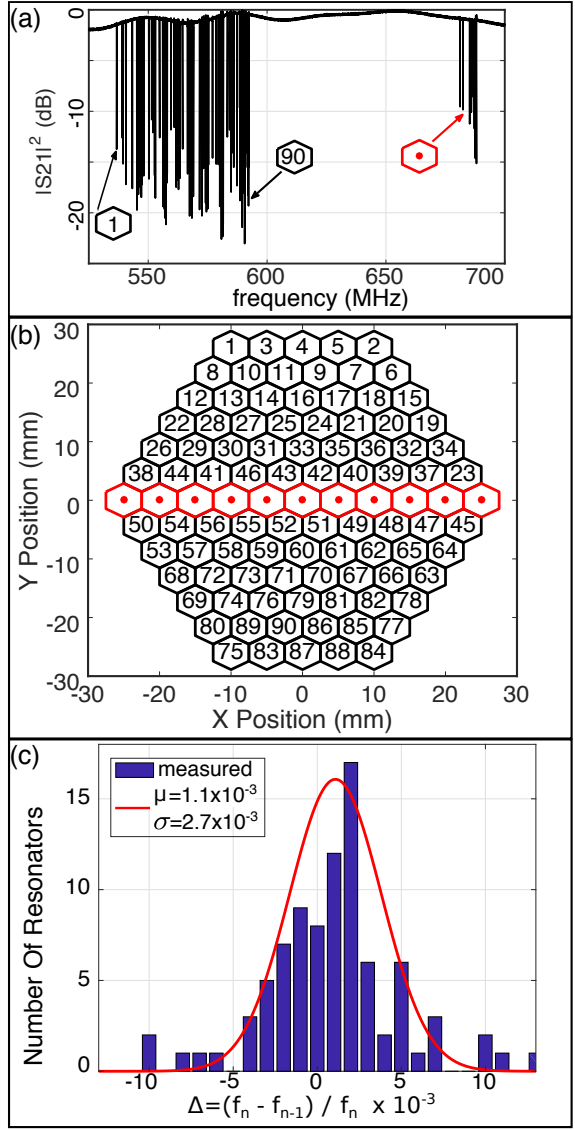}
      \caption{Test wafer frequency uniformity results.
      (a) S$_{21}$ trace of the array showing two distinct groups.  
      Resonators spanning 535-595~MHz correspond to group 1: an array with designed fractional frequency spacing $\Delta~=10^{-3}$ that spans the entire size of the wafer.  
      Resonators between 680-686~MHz correspond to group 2: 11 identically designed resonators in the center row.  
      The $1.1 \times 10^{-2}$ fractional frequency spread indicates the level of wafer-scale uniformity.  
      (b) Resonant frequency to physical resonator map.    
      Numbers correspond to frequency placement in ascending order.  
      The wafer was designed to follow a serpentine pattern starting from the top left.  
      The observed frequency shuffling is expected given that the designed $\Delta$ is a factor of 10 smaller than the observed frequency scatter in the group 2 resonators, which are displayed as red hexagons.
      (c) Measured fractional frequency spacing ($\Delta$) between spatially adjacent resonators.  
      Local deviations in $\Delta$ are approximately Gaussian distributed with $\sigma = 2.7 \times 10^{-3}$ and median value $\mu = 1.1 \times 10^{-3}$. }
  \label{Fig:Wafer0}
\end{figure}

\figrefstartsentence{Fig:Wafer0}(a) shows microwave transmission of the test wafer.  
The two design groups of resonators are easily identifiable.  
The resonator group spanning 535-595~MHz contains all 90 resonators of group 1.  
The higher frequency group (680-686~MHz) is group 2, the identically designed resonators of the central row.  

As motivated in the introduction, we are mainly concerned with the scatter of the fractional frequency spacing $\Delta$, as this impacts frequency collisions and array yield.  
For the resonators of group 2, we find that the maximum variation in $\Delta$ is $1.1 \times 10^{-2}$.  

Since the design $\Delta$ of the group 1 resonators is smaller than the observed scatter of $\Delta$ in group 2, we might expect shuffling of the frequency ordering of resonators in group 1 with respect to the designed serpentine pattern.  
By use of the recently developed LED pixel mapper \cite{LEDMapper}, we successfully mapped the measured resonator frequencies to the physical resonators.  
Indeed we observe frequency shuffling as shown in the numbered pixel layout of \figref{Fig:Wafer0}(b).  

Previous measurements of the group 1 resonators in this wafer found a relative frequency shift between resonators of order $\sim\;$1.5\% with a mostly radial dependence\cite{LEDMapper}, similar to the spread in group 2 resonators.  \figrefstartsentence{Fig:Wafer0}(c) plots the measured frequency separation between adjacent resonators, $\Delta = f_{n} - f_{n-1} / f_{n}$.  The scatter in $\Delta$ between spatially adjacent resonators is approximately Gaussian with $\sigma = 2.7 \times 10^{-3}$ and a median value $\mu = 1.1 \times 10^{-3}$. 

The radial dependence can largely be explained by variation in material parameters of the TiN/Ti/TiN superconducting film and ICP-RIE etch variation across the wafer. 
Variations in $T_{c}$ and $R_{s}$ lead to changes in kinetic inductance, as suggested by the Mattis-Bardeen thin film model where $L_{s} \propto R_{s} / T_c$ in the BCS limit\cite{MattisBardeen}.
Variation in the ICP-RIE over-etch into silicon ($z$) leads to changes in capacitance.  
Simulations show that the effective capacitance change of a unity width-to-separation ($\text{w}:\text{s}$) ratio IDC, with relative permittivity $\epsilon_{r} = 11.7$, is $\delta C / C \approx -1.34  \; \delta z / \text{w}$.
Together, the expected fractional frequency shift of a resonator from nominal values due to variation in these three physical parameters is
\begin{align}
    \label{eq:ExpectedFracShift}
    \frac{\delta f}{f} \approx  \frac{1}{2} \left(-\alpha \frac{\delta R_{s}}{R_{s}}  + \alpha \frac{\delta T_{c}}{T_{c}} +  1.34  \frac{\delta z}{\text{w}}\right).
\end{align}
The ratio of the kinetic inductance to total circuit inductance is $\alpha~=~0.82$, and $\text{w} = 2 \; \mu\text{m}$.  
\begin{figure*}[ht!]
  \centering
    \includegraphics[width=\textwidth]{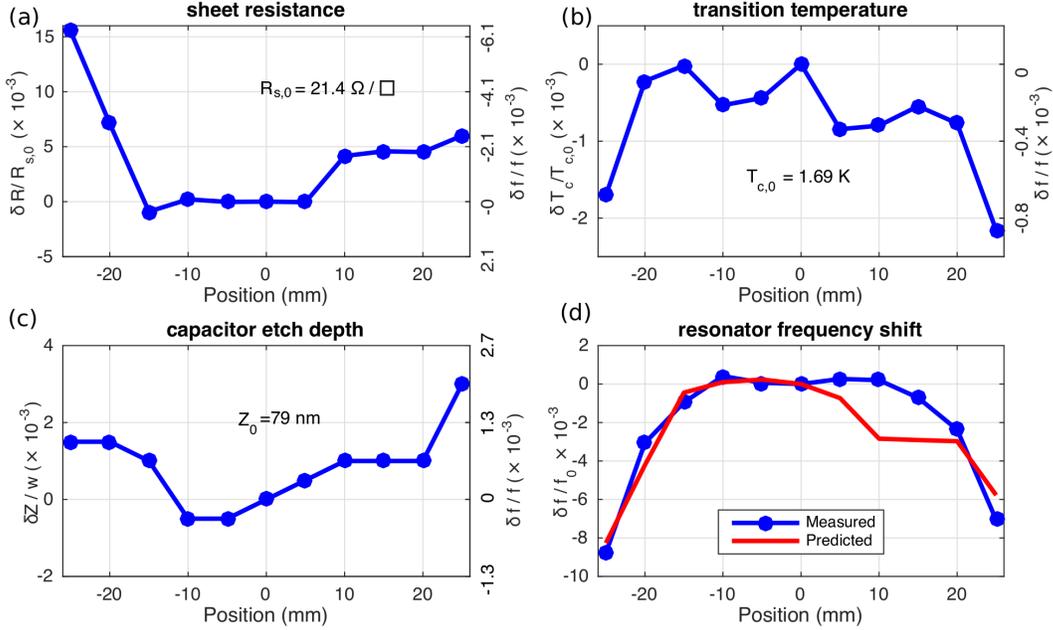}
      \caption{Variation of material parameters on test wafer versus radial position.  
      Fractional variation of (a) sheet resistance, (b) superconducting transition temperature, and (c) capacitor etch depth from the 11 identically designed center row MKIDs on the test wafer.  
      Fractional variations are normalized to the center resonator at $0 \; \text{mm}$.   
      The right axis in (a)-(c) shows the fractional frequency shift of the individual parameter predicted using the appropriate term in \eqnref{eq:ExpectedFracShift}.  
      (d) The measured fractional frequency shift is in good agreement with a model that utilizes \eqnref{eq:ExpectedFracShift} and the measured parameters in (a)-(c).}
    \label{Fig:SingleRow}
\end{figure*}

Measurements as a function of radial position of these three parameters for the group 2 resonators are presented in \figref{Fig:SingleRow}(a)-(c).  
The measurements suggest a maximum fractional frequency deviation of $<10^{-2}$ from each component in isolation.  The largest deviations are sourced from $R_{s}$, but all three components contribute fractional frequency deviations $> 10^{-3}$.

\figref{Fig:SingleRow}(d) shows that the measured frequency deviations of the group 2 resonators are in good agreement with the model, which assumes \eqnref{eq:ExpectedFracShift} and uses the measured film parameters of each sample.  
However, the agreement is not good enough to use when designing an array.  
Furthermore, we should not expect the exact radial dependence to hold wafer to wafer.

We conclude that the dominant source of frequency deviation is due to material parameters that vary smoothly across the wafer.  
Local variations in $\Delta$ are much smaller, at the level of $\sim10^{-3}$.  
This suggests a clear path to decrease collision rates: spatially neighboring resonators ought to be close in frequency space.
Using this approach, global variations will not strongly affect the yield.  Instead it will be determined largely by the scatter in $\Delta$ between adjacent resonators shown in \figref{Fig:Wafer0}(c).

\section{Design and characterization of the BLAST-TNG 250~$\mu\text{m}$ Focal Plane}

\begin{figure}
  \centering
    \includegraphics{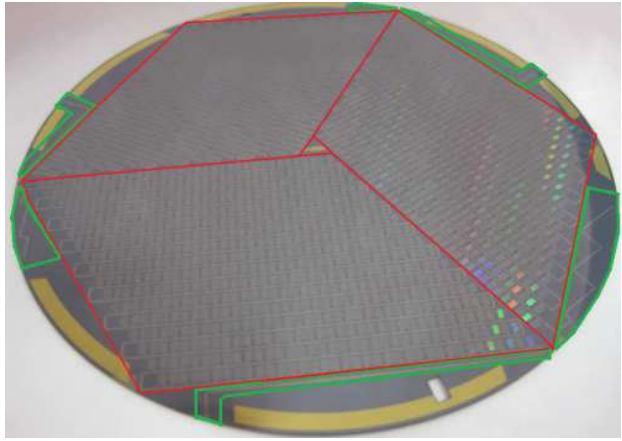}
      \caption{BLAST-TNG 250$\;$\micron\ MKID array.  
      The 93~mm diameter, 918 pixel (1836 MKID) array is composed of three copies of a rhombus (outlined in red) that are lithographically defined with the tile-and-trim method.  
      The 612 MKIDs in each rhombus couple to a microstrip feedline, which routes to the edge of the wafer (green outline) for interconnects.  
      Gold sections on the perimeter are used for array heat-sinking.}
\label{Fig:BLAST250_C}
\end{figure}

We applied the tile-and-trim fabrication method to the production of the BLAST-TNG 250~\micron\ array, shown in \figref{Fig:BLAST250_C}.  
BLAST-TNG is a balloon-borne submillimeter polarimeter that will image dusty star-forming regions by use of three monolithic MKID arrays, one for each waveband centered at $250 \; \mu\text{m}$, $350 \; \mu\text{m}$, and $500 \; \mu\text{m}$.  
The 250~\micron\ array consists of 918 polarization sensitive pixels (1836 MKIDs) that fill a 93~mm focal plane and use a 2.5~mm detector pitch.    
The focal plane is divided into three identical rhombii.  
For the lithography of the array, the stepper runs through one rhombus applying the tile-and-trim program.  
The wafer is then rotated 120$^\circ$, and the same program runs to produce the next rhombus.  One additional lithography step and deep etch on the back side creates an optical quarter-wavelength thick substrate below the detector absorbers, and subsequent metalization forms a backshort.
Each rhombus has a pair of microwave ports coupled to 306 pixels (612 resonators), which are intended to be coupled to one room temperature readout board that has 512~MHz of digitization bandwidth \cite{gordon2016}.  
Therefore, we design $\Delta = 1.02 \times 10^{-3}$ and layout the resonators such that spatial neighbors are closely spaced in resonant frequency, taking advantage of the expected high local uniformity that was observed in the test wafer. 

\begin{figure*}
  \centering
    \includegraphics[width=\textwidth]{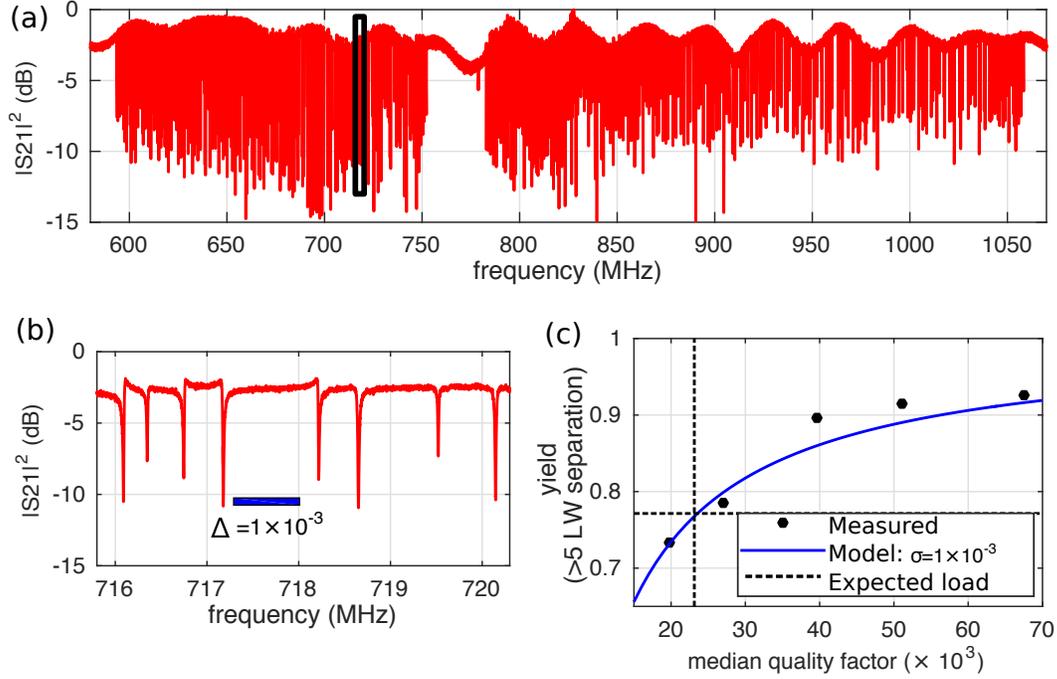}
      \caption{BLAST-TNG 250$\;$\micron\ frequency survey and yield.
      (a) $S_{21}$ of a single rhombus on the array operated at $T_b$~=~275 mK.  
      Lower and upper frequency groups correspond to the X and Y-Polarization-sensitive MKIDs, respectively.  
      (b) Zoom-in of the boxed region in (a). 
      The blue scale bar illustrates the design frequency spacing.
      (c) Yield (the fraction of resonators with $\chi > 5$ defined in \eqnref{Eq:TheorySurvive} as a function of median quality factor, which we vary by changing $T_{b}$.  
      The model (solid blue), produced using \eqnref{Eq:TheorySurvive} and $\sigma~=~1\times10^{-3}$, qualitatively matches the data.
      The dashed black line shows that the array yield is 77\% when operated in the expected conditions of the experiment ($T_b$~=~275~mK and 15~pW photon load, which produces $Q$~=~23$\times10^{3}$).
    }
    \label{Fig:BLAST250_S21_Zoom_Yield}
\end{figure*}

\figrefstartsentence{Fig:BLAST250_S21_Zoom_Yield} shows the frequency survey results of one rhombus of the 250~\micron\ array when operated in a dark enclosure.   
605 resonators are identified between 590~MHz and 1.055~GHz, indicating a physical detector yield of 99\%.  
The two polarization bands are separated by a $\sim\;$10~MHz gap centered at 775~MHz.  

The detector yield is a function of operating conditions because both the photon load $P_o$ and the operating temperature $T_b$ determine the resonator quality factor $Q$.    
Operating in a dark environment at $T_b$~=~275~mK, as is shown in \figref{Fig:BLAST250_S21_Zoom_Yield}(a), the measured detector yield ($\chi > 5$) is 89\%. 
By varying $T_b$, we determine the yield as a function of median quality factor (\figref{Fig:BLAST250_S21_Zoom_Yield}(c)).  
Comparison to the model of \eqnref{Eq:TheorySurvive} indicates an effective $\sigma \sim 10^{-3}$, similar to what was observed in the test wafer.   

The expected operating conditions of the 250~\micron\ array during observations are $T_b$~=~275~mK and $P_o$~=~15~pW.  
By use of single pixel measurements operated under these conditions, we determine a median $Q~=~23\times10^{3}$, and thus we expect 77\% detector yield under realistic observing conditions.

\section{Conclusion}

We have presented the tile-and-trim method for the fabrication of large-scale MKID arrays with the targeted goal of enabling high multiplexing factors.  
The technique has already been adapted by other processes.  
With its use in the fabrication of microwave SQUID multiplexers a maximum deviation in $\Delta$ of $2.9 \times 10^{-4}$ was demonstrated\cite{Dober2017}, which enables multiplexing factors of 2000.  
With a base pixel design in hand, an array is quick to layout and fabricate, and is easily reconfigurable to different substrate sizes.  
The method readily applies to fabrication on 150~mm wafers.
The same photolithography reticle set may be used for both prototyping devices and production arrays.

Through the characterization of a 101-pixel MKID test wafer, we find the maximum peak-to-peak frequency spacing deviation is 1.5\%, and is attributed to material properties that vary smoothly across the wafer.  The local scatter is approximately a factor of 5 smaller.

We exploited this finding to increase the yield of the BLAST-TNG 250~\micron\ production wafer by placing resonators close in frequency space also close in physical space.  
The yield measurements of this 1836 MKID array are consistent with an $rms$ scatter in $\Delta$ of $\sim 1 \times 10^{-3}$.

The yield is 77\% when operated in the conditions expected during astronomical observations.  
This value is consistent with our model of frequency collisions and is comparable to the yield achieved with other low temperature detector technologies.  

Further yield improvements may be gained by use of post-measurement lithographic correction \cite{2017MKIDFreqTrimming}.
Its use has demonstrated Gaussian distributed frequency spacing with $\sigma \approx 3.5 \times 10^{-4}$, a factor of 3 improvement. 
Applying this to the BLAST-TNG 250~\micron\ wafer, an array yield of 97\% under expected operating conditions is achievable.

This paper is a contribution of NIST and not subject to copyright.

\bibliography{MKIDBibliography} 
\bibliographystyle{ieeetr}

\end{document}